\documentclass[12pt,aps,tightenlines,amsmath,amssymb,nofootinbib,superscriptaddress]{revtex4}
\usepackage{amsmath,amsthm,graphicx,amssymb,verbatim,listings}
\usepackage[usenames, dvipsnames]{color}
\usepackage[ pdftex, plainpages = false, pdfpagelabels,
                 pdfpagelayout = useoutlines,
                 bookmarks,
                 bookmarksopen = true,
                 bookmarksnumbered = true,
                 breaklinks = true,
                 linktocpage=all,
                 pagebackref=false,
                 colorlinks = true,
                 linkcolor = BrickRed,
                 urlcolor  = blue,
                 citecolor = BrickRed,
                 anchorcolor = green,
                 hyperindex = true,
                 hyperfigures
                 ]{hyperref}
\usepackage{xifthen} 

\newcommand{\tr}{\hbox{tr}}

\newcommand{\arxiv}[2][]{\ifthenelse{\isempty{#1}}{\href{http://arxiv.org/abs/#2}{{\tt arXiv:\allowbreak{}#2}}} {\href{http://arxiv.org/abs/#2}{{\tt arXiv:\allowbreak{}#2 [#1]}}}}
\newcommand{\pirsa}[1]{\href{http://pirsa.org/#1/}{{\tt http://pirsa.org/#1}}}

\begin{document}
\title{Some Negative Remarks on Operational Approaches \\ to Quantum Theory}

\author{Christopher A. Fuchs}
\affiliation{Raytheon BBN Technologies, 10 Moulton Street, Cambridge MA 02138, USA}
\affiliation{(\href{http://www.physics.umb.edu/Research/QBism/}{Now in the Physics Department, University of Massachusetts Boston, Boston MA 02125, USA})}
\author{Blake C. Stacey}
\affiliation{Martin A. Fisher School of Physics, Brandeis University, Waltham MA 02453, USA\bigskip}
\affiliation{(\href{http://www.physics.umb.edu/Research/QBism/}{Now in the Physics Department, University of Massachusetts Boston, Boston MA 02125, USA})}

\begin{abstract}
Over the last 10 years there has been an explosion of ``operational
reconstructions'' of quantum theory. This is great stuff: For, through
it, we come to see the myriad ways in which the quantum formalism can
be chopped into primitives and, through clever toil, brought back
together to form a smooth whole. An image comes to mind of a
brain-teaser puzzle, all sliding and interlocking pieces. There is no
doubt that this is invaluable work, particularly for our understanding
of the intricate connections between so many quantum information
protocols. But to me, it seems to miss the mark for an ultimate
understanding of quantum theory; I am left hungry. I still want to
know what strange property of matter forces this formalism upon our
information accounting. To play on something Einstein once wrote to
Max Born, ``The quantum reconstructions are certainly imposing. But an
inner voice tells me that they are not yet the real thing. The
reconstructions say a lot, but do not really bring us any closer to
the secret of the `old one'.'' In this talk, I hope to expand on these
points and convey some sense of why I am fascinated with the problem
of the symmetric informationally complete POVMs to an extent greater
than axiomatic reconstructions.
\end{abstract}

\maketitle

\noindent {\bf The following is a lightly edited transcript of a
  talk by one of us (CAF) at the workshop on Conceptual Foundations and Foils
  for Quantum Information Processing, Perimeter Institute for Theoretical Physics,
  10 May 2011.  A full video is available at
  \pirsa{11050055}.}


\bigskip

I always like to start with a joke, but due to the Hollywood special effects of Charles H. Bennett, I've got
something to share with you.  I'll let Charles tell the joke.  Many years ago, Asher Peres and I wrote an article called ``Quantum Theory Needs No `Interpretation'\,''~\cite{Fuchs00}, or as Charles would have it:
\begin{center}
\includegraphics[width=12cm]{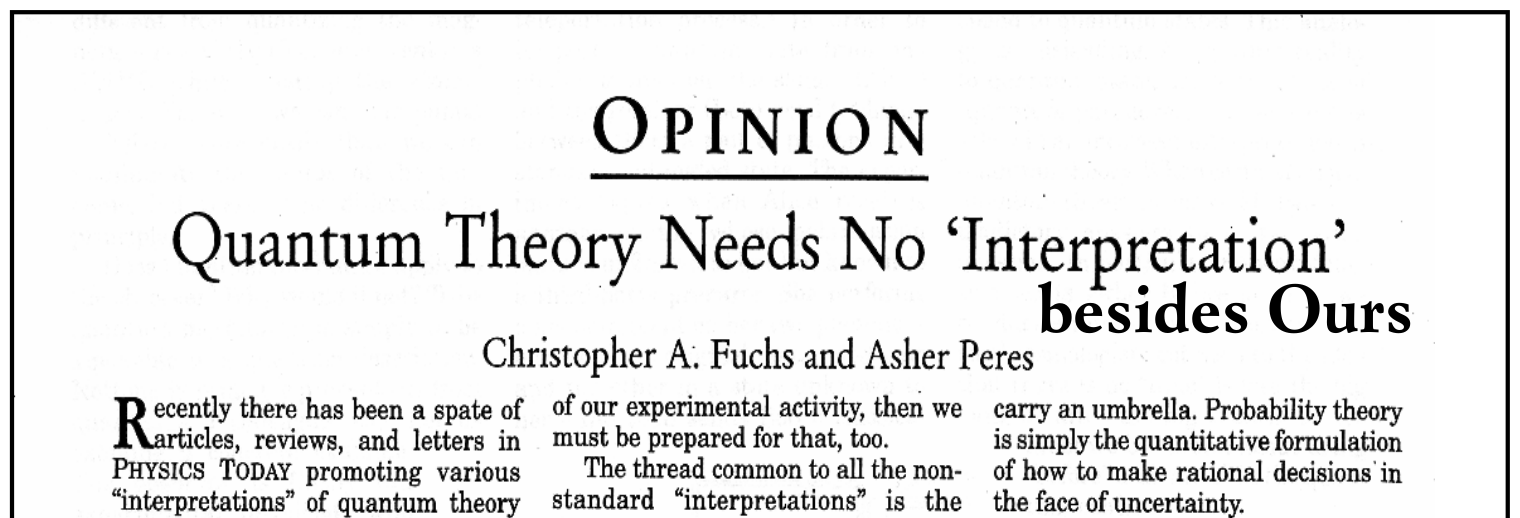}
\end{center}
But the paper ended with these words!
\begin{quote}
All this said, we would be the last to claim that the foundations of quantum theory are not worth further scrutiny.  For instance, it is interesting to search for minimal sets of {\it physical\/} assumptions that give rise to the theory.
\end{quote}

So, some negative remarks on operational approaches!  (I'm awfully loud, aren't I?)  [Someone in the audience: ``You always are!'']   I've been happy
to see that this old slide of mine, with the traditional axioms of
quantum theory, has gotten such airplay this week.
\begin{center}
\includegraphics[width=9cm]{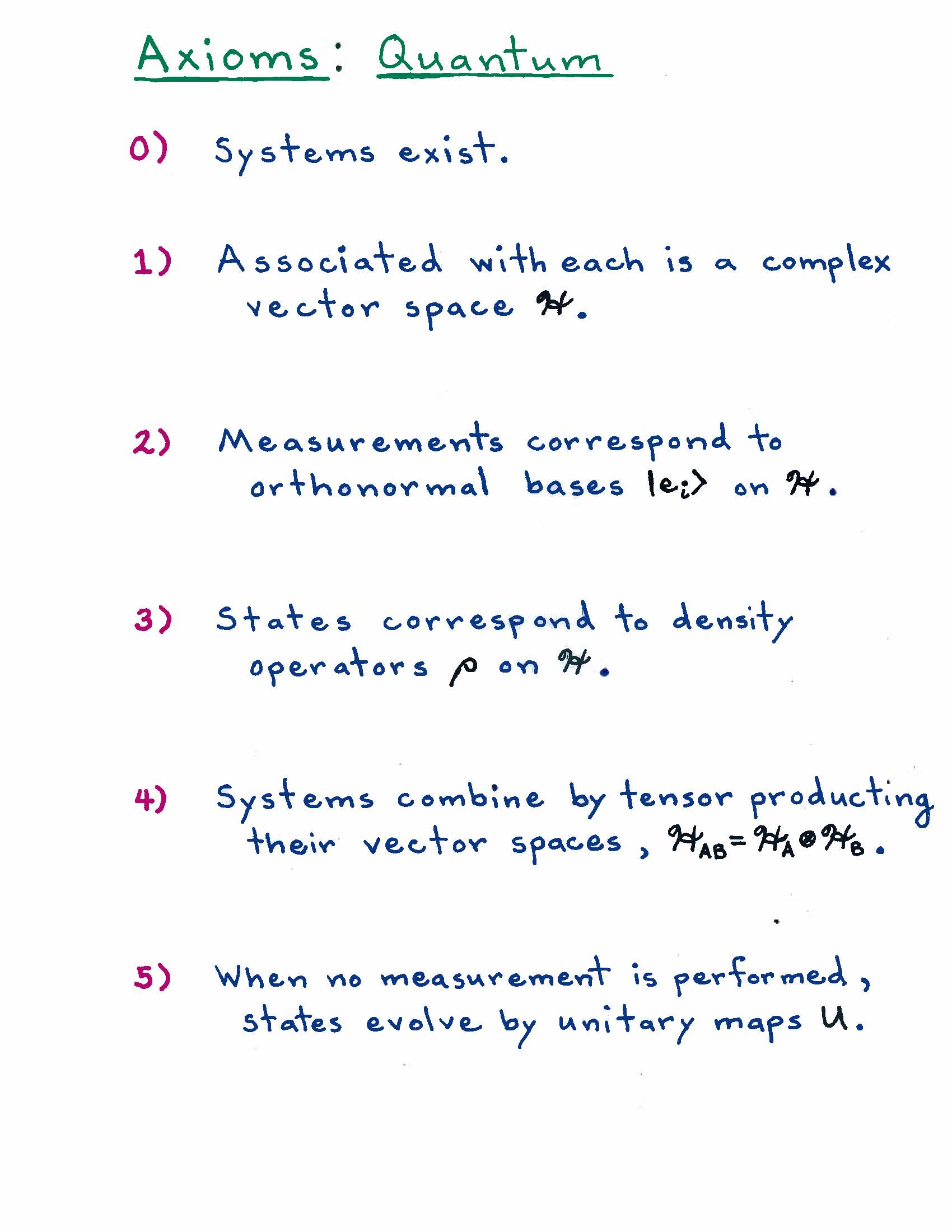}
\end{center}
\vspace{-1.5cm} I pulled it out of
the bin and I haven't used it in quite a long time.  In fact, I think
it was around 12 years ago, at some meeting in Maryland when I first
put the thing up.  At the time, it was mostly as an excuse to make a
joke at Max Tegmark's expense, who was going around to meetings taking polls of
which interpretation was the most popular at which meeting.
\vspace{-0.5cm}\begin{center}
\includegraphics[width=6.5cm]{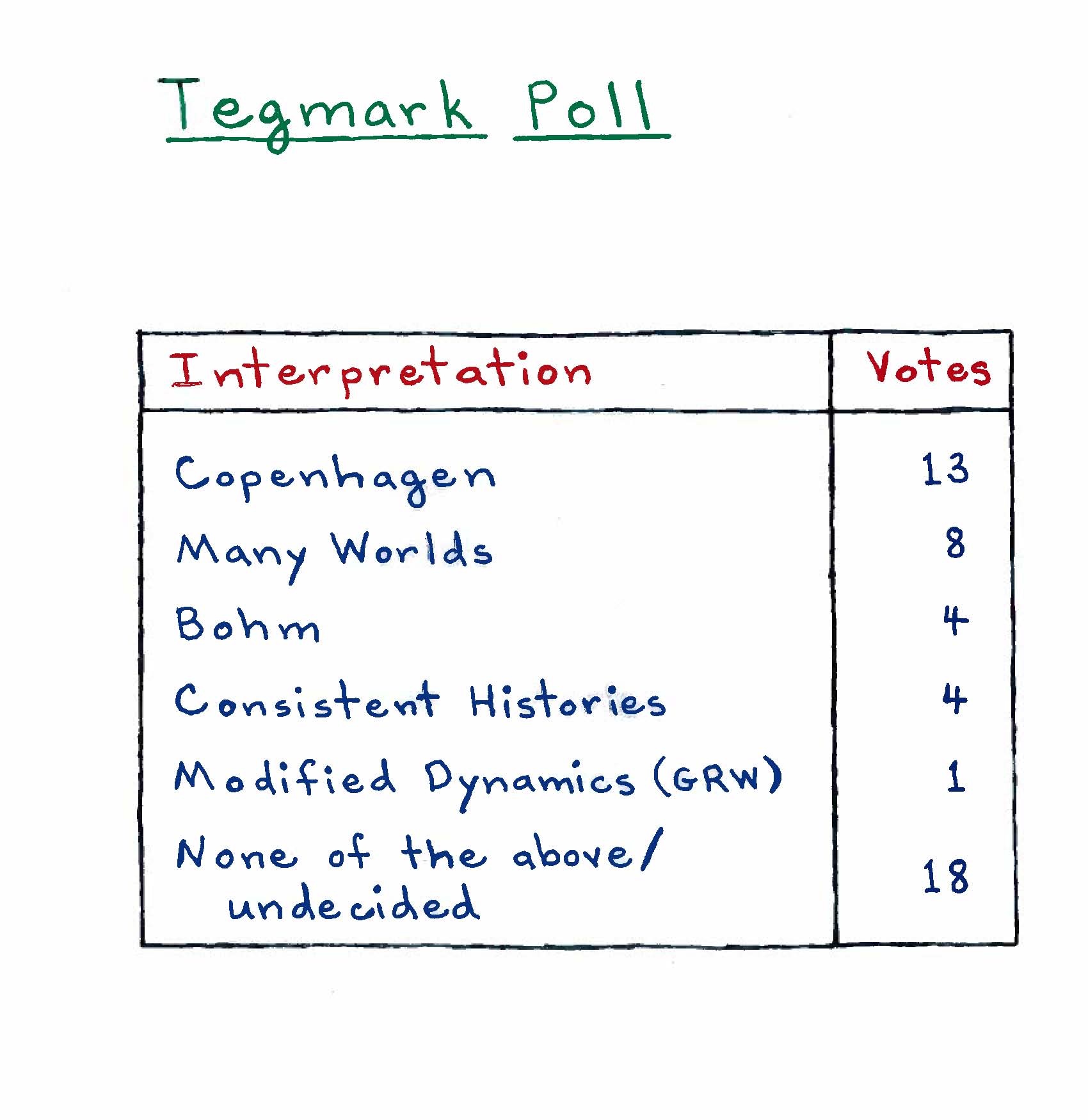}
\end{center}
\vspace*{-0.5cm} And there was a hidden agenda \ldots\ well it wasn't so hidden \ldots\ it was a pretty obvious agenda that he wanted to see Many Worlds climb in
numbers as time went by, so that he could say that the Many Worlds
Interpretation was the eminently reasonable one, and that's decided by
democratic vote.

All that caused me to reflect upon what our mission \ldots\ well, at the time I didn't call
myself quantum foundations, and I probably won't be calling myself
quantum foundations after this meeting \ldots\ it caused me to reflect upon
what exactly we needed to do as a community to get in a position that
we would disband.  That there wouldn't be any more quantum
foundational meetings---how could we make this sort of stale debate on
interpretation, where numbers fluctuated from year to year but
otherwise there was no great progress---how might we make that end?
And the thing that struck me was that this slide actually said it all.
This is one of our great physical theories, one of the great theories of physics, one of the two or three
great theories of physics, and look how it's posed!

Associated with each system is a complex vector space.  Vectors, tensor products, all of these things. Compare that to one of our other great physical theories, special
relativity.  One could make the statement of it in terms of some very
crisp and clear physical principles:  The speed of light is constant
in all inertial frames, and the laws of physics are the same in all
inertial frames.  And it struck me that if we couldn't take the
structure of quantum theory and change it from this very overt
mathematical speak---something that didn't look to have much physical
content at all, in a way that anyone could identify with some kind of
physical principle---if we couldn't turn that into something like
this, then the debate would go on forever and ever.  And it seemed
like a worthwhile exercise to try to reduce the mathematical structure
of quantum mechanics to some crisp physical statements.

Now the reason I went in a direction like this, where I said, ``What
would be a good methodology for doing that?,''\ and I landed upon
going to each and every one of the axioms and trying to give it an
information-theoretic reason {\it if possible}, was that by that time I had
already become pretty convinced that most of the structure of quantum
theory was about information~\cite{Fuchs01}.  I had convinced myself
that quantum states represented information of some sort, or Bayesian
degrees of belief, or some might say knowledge\ldots But the question
on my mind was, ``How much of quantum theory is about information?''
Just because some parts of it are about information, it didn't mean
that \emph{all\/} of it had to be about information.  And I threw my
money on the idea that there would be something about quantum theory
that was \emph{not\/} information-theoretic.

And so, when I wrote the paper~\cite{Fuchs02} that I took this little
image from, I wrote of all these axioms, ``Give an
information-theoretic reason {\it if possible}!''
\begin{quotation}
\noindent The distillate that remains---the piece of quantum theory
with no information theoretic significance---will be our first
unadorned glimpse of ``quantum reality.''  Far from being the end of
the journey, placing this conception of nature in open view will be
the start of a great adventure.
\end{quotation}

So that was really what I was seeking: to try and tear away all the
underbrush that was about information, and find the one piece, or some
small number of pieces, of quantum theory that were actually
statements about the world independent of information-processing
agents, independent of measuring observers and so forth.  What I was
really after was, I wanted to know what made quantum systems go?  What
made them interesting and behave in some peculiar way?  Could we
pinpoint that one thing on a principle that wasn't written in
information-theoretic terms or operational terms in any way?

So, some time has passed, and there's been all of this fantastic work!
And that's why I'm at this conference, because this has surprised me:
the number of people that really threw their hearts and souls and have made
all this great progress that we've seen here.  We've seen some of
these axiom systems---here's the one of Chiribella, D'Ariano and
Perinotti~\cite{Chiribella11, Brukner11}, their five axioms and one
postulate, where they lay down principles of causality, perfect
distinguishability, ideal compression, local distinguishability, pure
conditioning and purification.  And out of that, all written in
English, no mathematical equations---I like that!---one pulls together
the mathematical structure of quantum theory, {\it after a lot of work}.
But they nail it.  And, particularly, one thing I like about this
system is the way they distinguish the purification postulate from
the other axioms.  This now approaches something that I think is along the
lines of trying to find a crisp physical principle.

\begin{center}
\includegraphics[width=14cm]{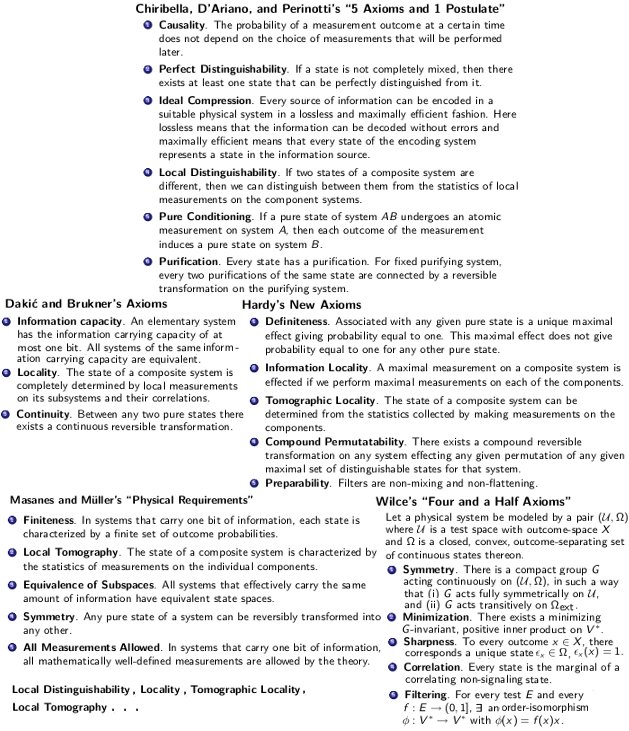}
\end{center}
I've put all you guys in alphabetical order, so not to offend anyone.
Daki\'c and Brukner's axioms: information capacity, locality,
continuity~\cite{Dakic09}.  (They haven't presented on this yet; I
guess you will later in the conference.)  Hardy's new axioms---we saw
his old axioms~\cite{Hardy01, Schack04} before, ten years ago, and his new
axioms this week---definiteness, information locality, tomographic
locality, compound permutability and preparability~\cite{Hardy11}.
Masanes and M\"uller's axioms, we've already seen this as well: finiteness, local tomography,
equivalence of subspaces, symmetry, all measurements are
allowed~\cite{Mueller12}.  And we saw Alex Wilce's this morning,
written in a little bit more mathematical language---there are some
equations in there!~\cite{Wilce09}

Well, when this whole business---or at least my discussions with
Gilles Brassard and Charlie Bennett and many of you started up---I
remember Charlie Bennett saying, ``How will you know when you get to
this physical distillate of quantum theory?  How will you know that
you've reached the end of the process?''  And he sort of jokingly
said, ``Will it be like pornography?''  You know, ``the only way you know it's
pornography is if you see it.''  And I think he was right!  Not in the
way he wanted to be.  But I think what I am having a difficulty with,
and what I want to try to express is that, of all the progress that's been
made, I haven't been able to look at these systems and see something
that stands out at me as the essential core of quantum mechanics.
Something that's written in physical, nonoperational,
noninformation-theoretic terms.  So, there's been great progress in putting
everything in operational terms, and making that very clear, and one can
do that.  Can one do the opposite thing?  Can one find some
nonoperational terms and have most of it in operational or
information-theoretic terms but perhaps not all of it?

\vspace{-0.3cm}\begin{center}
\includegraphics[width=7cm]{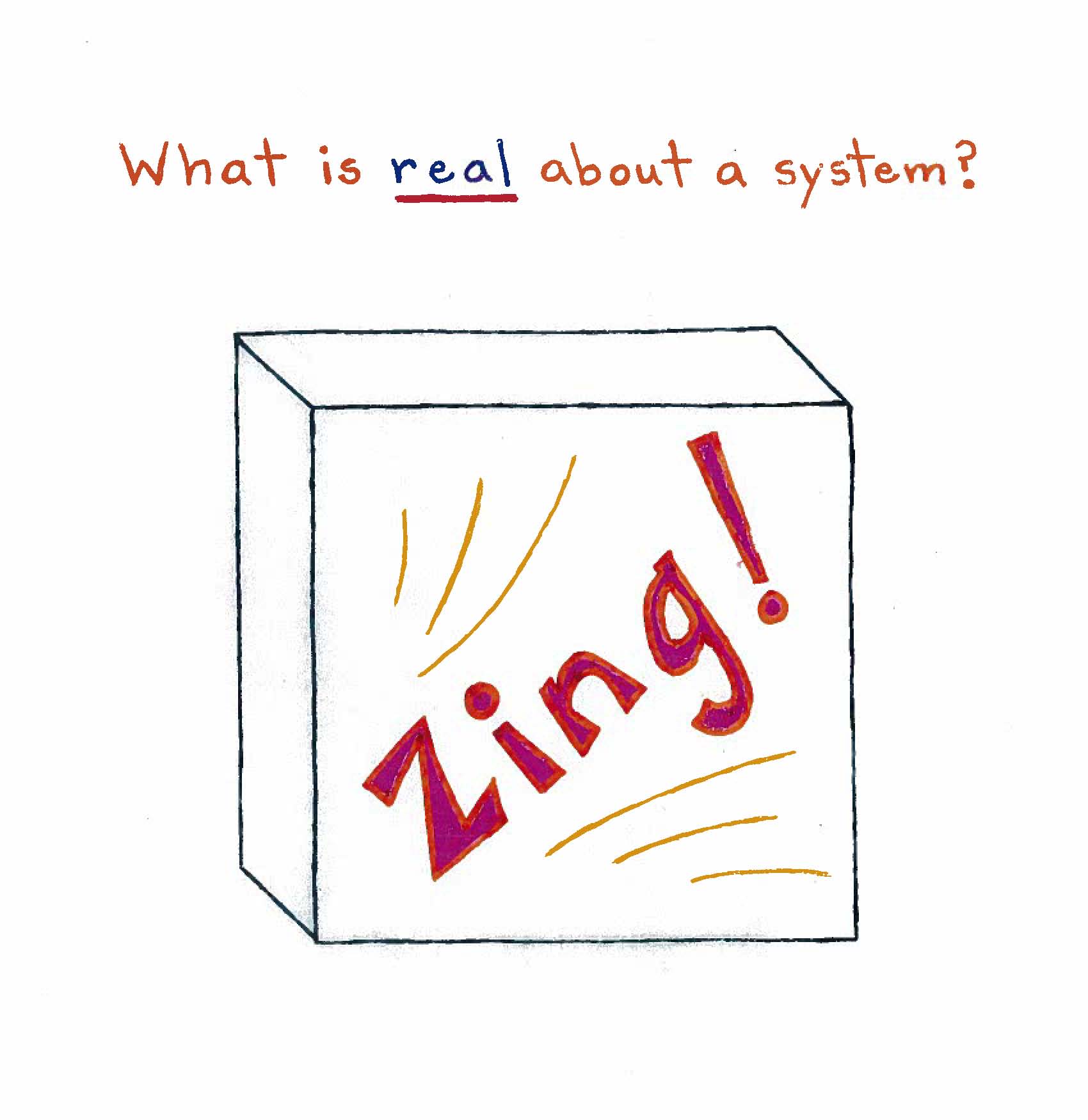}
\end{center}
\vspace{-0.8cm}

So, I look at these systems and I say, ``What is the distillate that's
left behind?  That I can say this is the Zing, this is the thing that makes quantum
systems go?''

And I don't see it.  So that's my basic point.

How would I know it if I saw it?  Back to Charlie's question.  I'm not
completely sure, but I think there's a distinction between the two
principles for relativity---which of course are girded up by some mathematics, as
Lucien Hardy made clear---and the ones that we've seen in the present
efforts.  And it's this: There's a certain amount of shock value in
seeing the two things side-by-side.  They're not just sort of inert
principles that are lying there and are equivalent to the structure of
the theory.  But instead, one of them says that the speed of light
should be constant in all inertial frames.  And if you're accustomed
to thinking the speed of light is this measurable quantity, how could
it possibly be the same in all frames?  Putting these two things together
caused a certain amount of shock.  So, I wonder whether quantum theory
can be written in terms where the physical principle is identified as
having a certain amount of shock value.

I guess another thing I'm trying to say is that of the existing axiomatic
systems, it doesn't seem, to me at least, that any have quite gone for
the jugular vein of quantum theory.  You know, when a werewolf attacks
a person he'll jump at this vein, and the person will bleed to death, and it'll all be over with.  I would
like to see an axiomatic system that goes for the weirdest part of
quantum theory of all.  For instance, most of the ones in the list
above are built on the idea of local tomography.  Is this a feature of
quantum theory that's really weird?  No, it doesn't seem to me that
it's a feature that's really weird.  And similarly with so many of the other ones.  Can we find some axiomatic system that really goes after the {\it weird\/} part of quantum theory?

Well, what is the weird part?  What is the toy that one might want to
go after for axiomatizing?

Is it nonlocality?  We hear this word all the time in quantum
information and quantum foundations.  My own sympathy, however, lies
with something that Albert Einstein said I believe in the '40s~\cite{Einstein48}.  It's in small print, I'll read it.  ``Einstein on Locality'':
\begin{quotation}
If one asks what is characteristic of the realm of physical ideas
independently of the quantum-theory, then above all the following
attracts our attention: the concepts of physics refer to a real
external world.
\end{quotation}
I'm okay with that!
\begin{quotation}
\noindent \ldots i.e., ideas are posited of things that claim a ``real
existence'' independent of the perceiving subject (bodies, fields,
etc.), \ldots.
Moreover, it is
characteristic of these physical things that they are conceived of as
being arranged in a space-time continuum. Further, it appears to be
essential for this arrangement of the things introduced in physics
that, at a specific time, these things claim an existence independent
of one another, insofar as these things ``lie in different parts of
space.''
\end{quotation}
He puts scare quotes around ``lie in different parts of space'' I
presume because he's meaning this is a tautology: we say things are in
different parts of space if they can't directly influence each other.
\begin{quotation}
\noindent Without such an assumption of the mutually independent
existence \ldots
\end{quotation}
This is the important part now.
\begin{quotation}
\noindent
Without such an assumption of the mutually independent
existence (the ``being-thus'') of spatially distant things, an
assumption which originates in everyday thought, physical thought in
the sense familiar to us would not be possible. Nor does one see how
physical laws could be formulated and tested without such a clean
separation.
\end{quotation}

I think this is maybe a debatable point in terms of detail, but I
think the idea is fundamentally sound.  If you first posit two
systems, and then you say, ``Oops! Made a mistake, they were really
one after all, because any one system can influence any other one,''
it would be hard to imagine how we come across the usual sort of
reasoning that we do.  So, my money is not on taking nonlocality as
the kind of shock-value principle I'm looking for.

Instead, I'm much more sympathetic to something Asher Peres would have said, or did say!, around 1978:
``Unperformed experiments have no results''~\cite{Peres78}.  If I were looking
for a shock-value principle to blame quantum mechanics on, my feeling
is that it's something along these lines.  What is really at issue
here?  It's the question of whether quantum measurements reveal some
pre-existing value for something that's unknown, or whether in some
sense they go toward creating that very value, from the process of
measurement.  A good way to see how to pose this in more technical
terms comes from---I guess I could call it a version of the Free Will
theorem~\cite{Conway06}, but it's much older than Conway and
Kochen~\cite{Caves07}---has to do with first looking at the EPR criterion of
reality~\cite{EPR}.  (The bracketed part is my addition.)
\begin{quotation}
\noindent If, without in any way disturbing a system one can [gather
  the information required to] predict with certainty (i.e., with
probability equal to unity) the value of a physical quantity, then
there exists an element of physical reality corresponding to this
physical quantity.
\end{quotation}
So, what's the content of that?  Is it right?

You can consider a little variation of the EPR thought experiment.
Let me take two qutrits and ascribe to them a maximally entangled
state, and let me assume locality, the principle that Einstein used in
the quote I took from him.  And now, consider making a measurement on
the left-hand particle in some basis---let's say this purple basis, or
alternatively this green basis.  I'm going to choose either the green
one or the purple one.  If I get, let's say, outcome number~2 on the
left side, then I can predict that if I were to make the measurement
on the right side, I would get outcome number~2.  (I've omitted a
transpose in drawing this picture.)  So, under the assumption of
locality, EPR would say, ``Aha!  It must be the case that there is an
element of reality on this side corresponding to outcome number~2 of
that measurement.  It's something inherent in that body.''  But
similarly, we could talk about the other measurement.  And if I were
to get outcome~2 for that measurement on one particle, I would predict
with certainty that I would get outcome~2 for that measurement on the
other particle.

\begin{center}
\includegraphics[width=7cm]{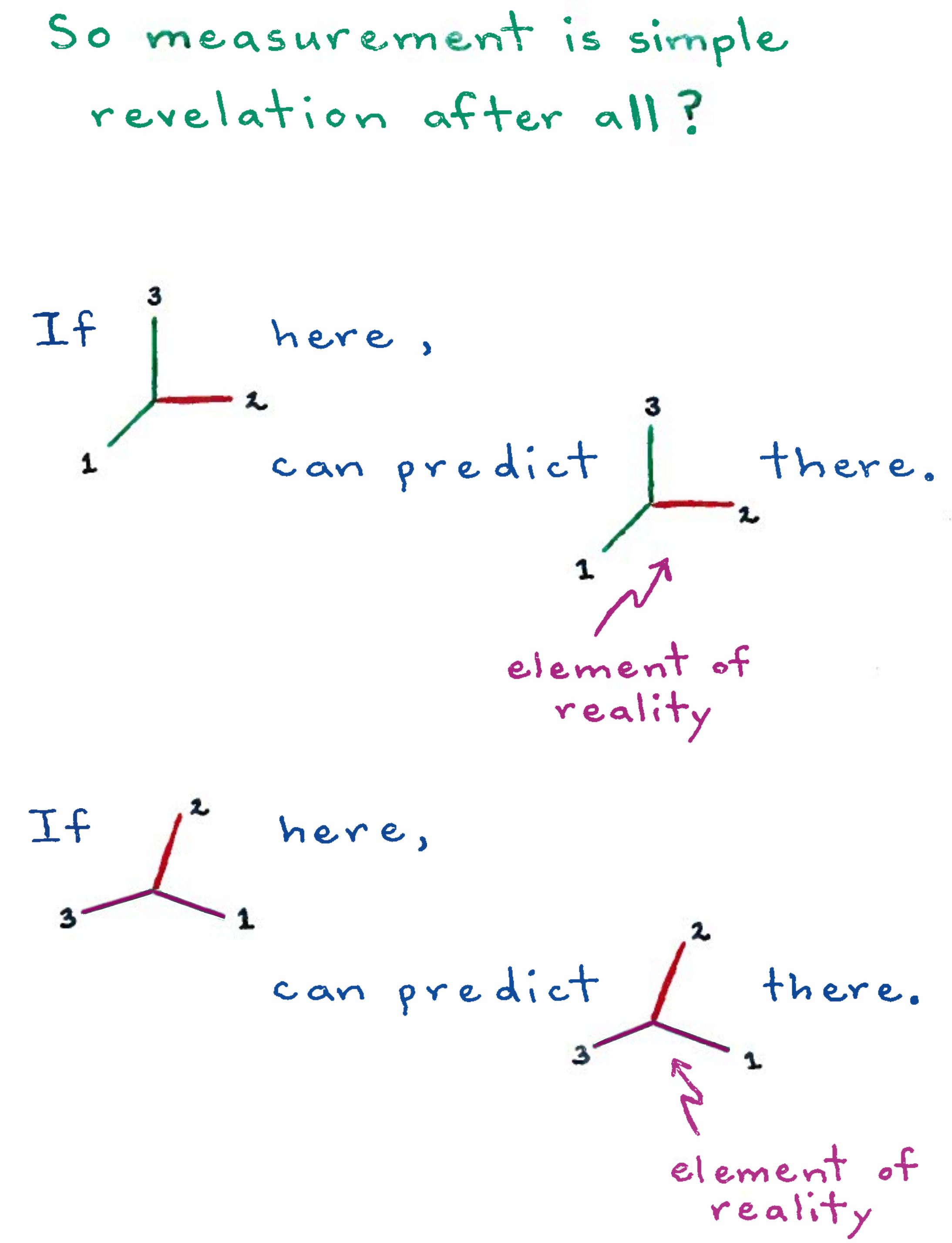}
\end{center}

We're talking about noncommuting variables here; by the EPR criterion
and locality, they would say it must be the case that there were
elements of reality associated with those noncommuting observables.
But we could make this more extreme and consider not just two
observables but a whole set of them, quite a lot of them, corresponding to one of the
Kochen--Specker constructions, for instance the one that Asher Peres
found~\cite{Peres91, PeresBook, Peres96}.  By taking a sufficient
number of orthogonal bases, and interlocking them in some interesting
way, we can construct a noncolorable set.  For each of these, we would
have said, by making a measurement here, I draw an inference about the
element of reality over here.  I can do it for one, I can do it for
another, I can do it for another \ldots\ I can't do it for all of them
without running into trouble!

\begin{center}
\includegraphics[width=9cm]{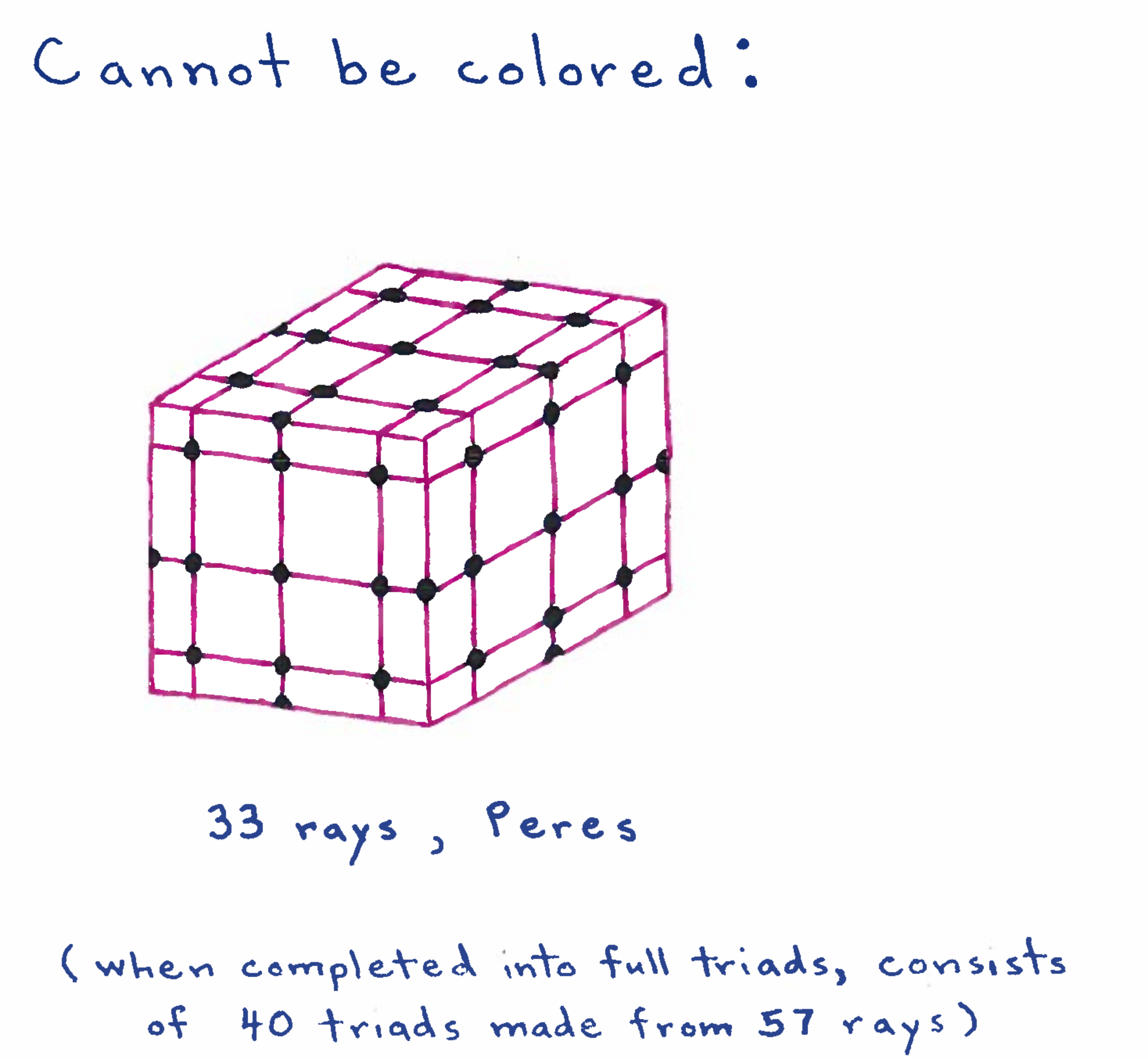}
\end{center}

I would say that the thing that this teaches me is that there's
something bankrupt not in locality, but bankrupt in the EPR criterion of
reality.  There's something wrong about that.  And that's what I'm
calling Peres's principle that unperformed experiments have no
outcomes.  Particularly, quantum probabilities are not probabilities
for some pre-existing reality that we're finding when we perform a
measurement, but instead for something that's produced in the process
of measurement.

So, what would excite me in an axiomatic approach to quantum
reconstruction?  I think it would be if someone could find a set of
axioms that pulled this idea to the forefront and really made it the
core of things.  Now I don't know how to do that!  At all!  But I have
a little toy approach that I've been playing with for some time that
pleases me at least.  And that is to recognize that these
considerations of Kochen--Specker, including this case where we have the locality assumption, rest on thinking about different
experiments in terms of each other.  We think about what happens if we
perform one experiment, we think about what happens if we were to
perform another one---and it tells us that contextuality is at the core of
our considerations.  What I would like as a goal is a way to push quantum
theory's specific form of contextuality all into one corner.  If I
could show that all of the phenomena of Kochen--Specker and Bell
inequality violations and everything else, the whole formal structure,
comes out of one corner to do with something with contextuality, I think I would be
pleased.

So, let me give you a progress report on that kind of idea~\cite{Fuchs11,
  Appleby11, Fuchs13}.

To give you the report, I have to make use of a little device I've
told so many of you about so many times: a particularly interesting
measuring device that I would like to elevate to a standard quantum
measurement.  It would be an informationally complete measurement.  In
other words, if I knew the statistics of the outcomes of this
measurement, I would be able to reconstruct the quantum state that
gave rise to them.  That is what I mean by this:  I'm getting rid of this symbol $\rho$ which
represents a quantum state and putting in its place a probability
distribution.  This probability distribution refers to the probabilities of the outcomes of this measurement up in the sky, and if this is an informationally complete measurement, we can completely reconstruct the quantum state $\rho$ from the probabilities.  This means that we can really cross out $\rho$ completely and use the probabilities instead.

Particularly, to make everything that I say have a pretty form, this
device should have the following properties.  Suppose you could find
$d^2$ rank-one projection operators, with this nice symmetry
condition:
\begin{equation}
\tr\, \Pi_i \Pi_j = \frac{1}{d+1},\ i \neq j.
\end{equation}
Namely, the Hilbert--Schmidt inner product between any two of the
projectors is equal to a constant value determined by the dimension.
If you can find $d^2$ projection operators satisfying this symmetry,
you can prove that they have to be linearly independent---they don't
have a choice!  Moreover, if you renormalize by~$1/d$, then they'll
sum up to the identity.  So, these form the elements of a POVM, and
they can be thought of as the outcomes of a quantum measurement.  We
call it a SIC measurement, where the acronym stands for ``Symmetric
Informationally Complete''~\cite{Fuchs13, Zauner99, Renes04,
  Bengtsson10, Scott10, Appleby2011b, Rosado11, Tabia12, Tabia13,
  Appleby13}.

Since this measurement is informationally complete, one can completely
determine the quantum state $\rho$ in terms of the probabilities given
by the Born Rule:
\begin{equation}
p(i) = \frac{1}{d} \tr (\rho\Pi_i).
\end{equation}
And because of this great symmetry, there's a lovely reconstruction
formula which says that the initial quantum state is just a linear
combination of the projection operators, where the expansion
coefficients are determined by the probabilities in a really simple
way:
\begin{equation}
\rho = \sum_i
 \left[(d+1)p(i) - \frac{1}{d}\right]\Pi_i.
\end{equation}
It's just a nice little affine transformation.

We can now explore what state space looks like in terms of
probabilities.  If we were talking about density operators, we could
say, ``We know what state space looks like: It's the set of positive
semidefinite matrices with trace one.''  But what does the state space look like in terms
of the probabilities themselves?

If you put an arbitrary probability distribution into the
reconstruction formula, you have to get a Hermitian matrix, because
this is a real combination of projection operators.  So you will always get a Hermitian matrix. But for certain
choices of the probabilities, you won't get a positive semidefinite
one.  What this tells you is if you want to specify the set of rank-one positive
semidefinite operators, all you have to add to the condition that you
have a Hermitian operator is that the trace of the operator squared
and the trace of the operator cubed both equal~1:
\begin{equation}
\rho^\dag = \rho,\ \tr\, \rho^2 = \tr\, \rho^3 = 1.
\end{equation}
And if you translate that into a condition on the probabilities, you
get two equations.  One says that the probabilities should lie on the
surface of a sphere of a certain radius:
\begin{equation}
\sum_i p(i)^2 = \frac{2}{d(d+1)}.
\end{equation}
And the other one says that they should satisfy a certain cubic condition
which isn't necessarily very pretty:
\begin{equation}
\sum_{jkl} c_{jkl} p(j)p(k)p(l) = \frac{d+7}{(d+1)^3},
\end{equation}
where
\begin{equation}
c_{jkl} = \hbox{Re}\,\tr (\Pi_j\Pi_k\Pi_l).
\end{equation}
But what you can glean from this is that the extreme points form the
intersection of two sets, one of which is a sphere and one of which is
some cubic curve.  So, the pure states can't be the whole sphere,
unless the latter condition is trivial: They're going to be some
subset of a sphere.

What does all this have to do with contextuality, and maybe a reconstruction
of quantum theory by some means associated with contextuality?  You can start to get a sense of
what that has to do with this from the following little thought experiment.
\begin{center}
\includegraphics[width=10cm]{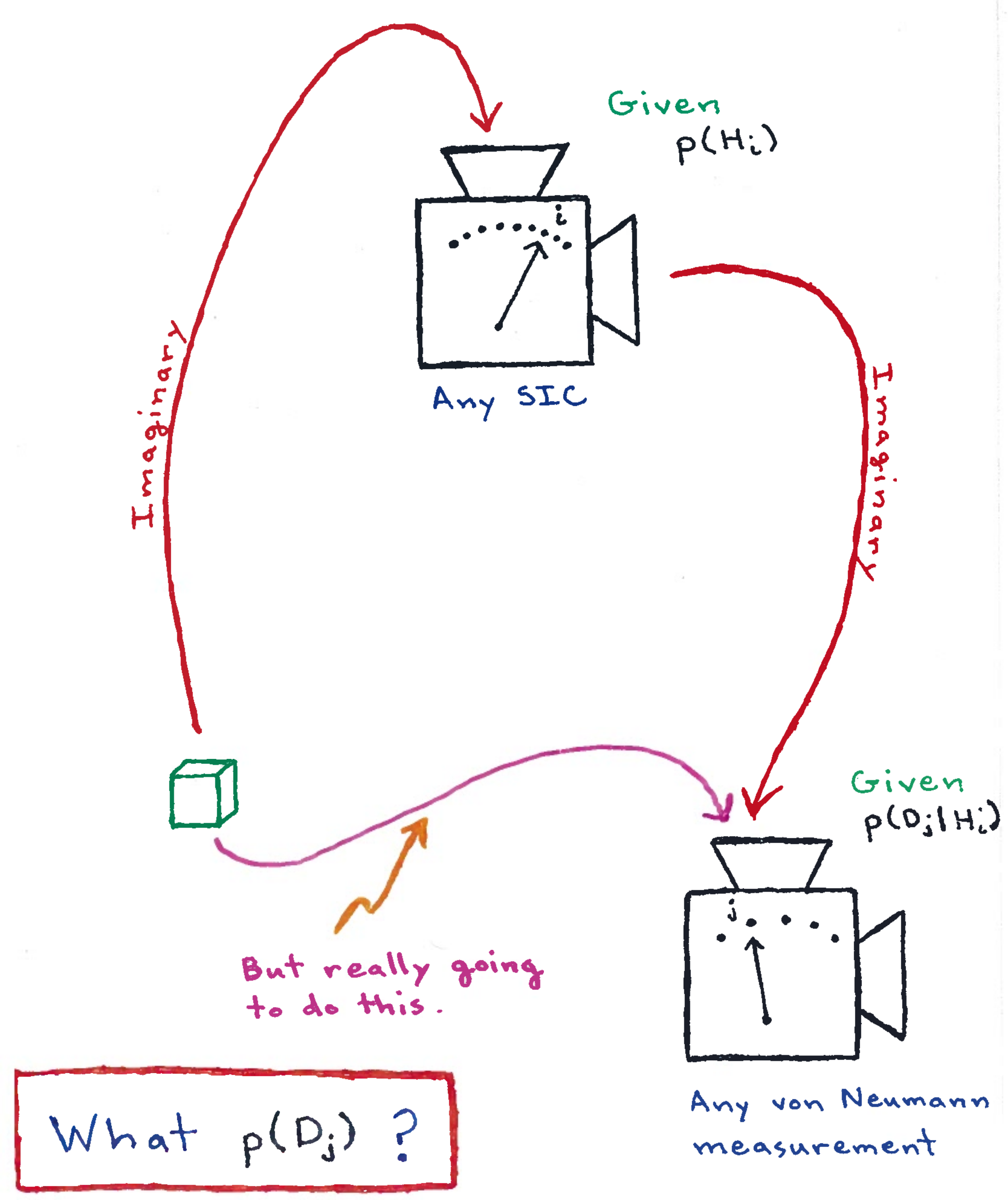}
\end{center}
Here's what I'm going to do.  I've got my quantum system, and I'm going to throw it into some
measuring device---I'll call it ``the measuring device on the
ground.''  Let's say it's a von Neumann measuring device.  So I'm going to throw this
quantum system into here, and when I do, it will generate an outcome. Let me call it $D_j$.  Here, I'm using the letter $D$ to evoke the idea of ``data''.  What I'd like to know is, what are the probabilities for the outcomes $D_j$?

$p(D_j)$---that's what I would like to calculate.  I'm going to throw the system in here; that's what I'm really going to do.  Throw it in there and get an outcome.
But suppose I only know the probabilities for the outcomes of this very special SIC
measurement, and I know the conditional probabilities for the outcomes
of the measurement on the ground in terms of the measurement in the
sky.  In other words, I have conditional probabilities $p(D_j|H_i)$.  Do I have
enough information to reconstruct the probability I really want?

Well, you might be able to reconstruct it, but you can't use regular probability
theory.  You can't just say that the probabilities for the bottom path
will be given by the classical formula to do with the conditional
probabilities for the upper path.
\begin{equation}
p(D_j) \neq \sum_i p(H_i) p(D_j|H_i),
\end{equation}
because in physics terms one of them is a coherent process, and the other one is an incoherent process.
If you speak like a probabilist (or like Leslie Ballentine, when he
was here and I made this slide for him) you might say there are extra
conditions.  The probability on the left-hand side refers to one
experiment, and the probabilities on the right-hand side refer to a
different experiment.  So there's an extra conditional in these expressions, and it's no wonder that you don't get equality in
the classical expression because there are extra conditions floating
about:
\begin{align}
p(D_j) &\hbox{ is really } p(D_j|C_1), \nonumber\\
p(H_i) &\hbox{ is really } p(H_i|C_2), \\
p(D_j|H_i) &\hbox{ is really } p(D_j|H_i,C_2).\nonumber
\end{align}
What is so interesting in this case, however, is that though all of that is true---it's true
you can't just use the classical formula for the upper path to
calculate the lower path---quantum mechanics nonetheless provides the
tools for making the calculation.  {\it It's just a different calculation.}

In usual ``physicist language,'' what is going on is that the probability I really want is one that I get by throwing my quantum system into the second measuring device via a coherent path.  There's no measurement on it; there's no decoherence.  Yet, the probabilities I am given are ones to do with this imaginary path.  An incoherent path.

In any case, the calculation is this:  You do the old classical
calculation---first form this, the usual Bayesian approach,
\begin{equation}
\sum_i p(H_i) p(D_j|H_i),
\end{equation}
and then you simply stretch the answer a little bit by a factor that
depends on the dimension, and renormalize:
\begin{equation}
p(D_j) = (d+1)\sum_i p(H_i) p(D_j|H_i) - 1.
\end{equation}
So whereas raw probabilistic considerations say there is an
inequality, or at least that there can be an inequality in this expression, quantum mechanics restores equality by
changing the formula a little bit.  Quantum mechanics adds something,
gives some extra structure to probability theory that just raw, plain probability
theory itself does not have.

Again, just to emphasize.  To get this answer: you just do the classical calculation, and then to get the Born rule
you simply modify the classical calculation ever so little!
\begin{align}
p(D_j) &= \tr(\rho \hat{D}_j) \nonumber\\
 &= (d+1) \sum_i p(H_i) p(D_j|H_i) - 1 \label{tricksterism}\\
 &= (d+1) [\hbox{classical calculation}] - 1.\nonumber
\end{align}

This causes me to wonder whether this is the kind of corner where I can
push all of contextuality.  The contextuality here is that I've got
the calculation for the probabilities directly for the ground path in
terms of the probabilities for the sky path, and the formula is
modified.  Can we take this diagram and this addition to probability
theory as a fundamental postulate of quantum mechanics?

Well, what is one wanting to get out of it?  The thing you'd like to get out of it is a specification of this convex set
that we were talking about earlier.  Can it be that this formula somehow
implies the convex set that we were talking about, where we had a
sphere intersecting a cubic curve?  It might imply some features, and
you can already see that it might because of the following: Suppose the term
under the sum here were a number near zero, or zero in fact.  Well, if the number were zero,
then we would have $0 - 1$, and that would lead to a negative number
on the left-hand side.  So we wouldn't actually have a probability on the left-hand side!
We might have something that you would call negative probability, but negative probability isn't probability.
Alternatively, suppose the sum were close to 1, or in fact 1 itself, then the
right-hand side would be $(d+1) - 1$.  That would be $d$, and that would be
larger than~1.  So again, this term under the sum can't be too large; otherwise, we'll end
up with something that is not a proper probability either.  So it gives some hope hope that making this formula hold will give us something like a
convex structure, and something along the right variety of one.

OK, so what I {\em really\/} want to do---that was all tricksterism, I don't really want to take this one statement [Eq.~(\ref{tricksterism})] as a postulate of quantum mechanics.
My starting point should be to consider the most general case,
not just where I have a von Neumann measurement on the ground, but
where I have any kind of measurement whatsoever on the ground.
You can prove that the rule gets a little modified when we consider
completely general positive operator valued measures on the ground: The thing that changes is that there is an extra conditional term underneath the sum:
\begin{equation}
\boxed{\rule{0cm}{0.9cm}\;\; q(j) = \sum_{i=1}^{d^2}
 \left[(d+1)p(i) - \frac{1}{d}\right]r(j|i).
\;\;}
\label{Rufus}
\end{equation}
So in the case where we have a von Neumann measurement, then all of the rightmost terms under the sum---they sum up to~1 in fact.  We thus get this little extra term that is outside of the sum [in Eq.~(\ref{tricksterism})].

I am going to change my notation slightly, because I have more equations to
show you.  I'm calling the quantum probability, the thing the Born
Rule calculates for us, $q(j)$; I'll call the prior probabilities for
the measurement in the sky $p(i)$; and I'll call the conditional
probabilities for the ground outcomes given the sky outcomes $r(j|i)$.

Here's what I want to show you.  Nearly the consistency of this equation alone [Eq.~(\ref{Rufus})]---when I say ``nearly,'' you have to take that a little bit with a little grain of salt, but for me, it's nearly---nearly the consistency of this equation alone implies a significant, nontrivial convex structure.
You don't just end up with a sphere, you don't end up with a cube, you
don't end up with any polytope---you end up with something that you
don't see discussed much in these circles, at least as far as I've
seen.

Here's the ``nearly.'' Here's a property I'm going to add to that
axiom.  It's a property that quantum mechanics has.

Suppose your initial state is the complete garbage state, $\rho =
\frac{1}{d}I$, and you actually follow the path in the sky---so in our diagram, we throw the garbage state up into the sky, it goes through the SIC-POVM, and it comes down to the ground for some new POVM, which I'll call $\{G_j\}$.  So those are the elements of it.  Suppose I look
at the outcome $j$ and try to make an inference to what happened up
in the sky.  Well, then I'd just use the normal Bayes' Rule to calculate the
probability for the outcome $i$ in the sky given the outcome $j$ on
the ground.  And it just works out to be this:
\begin{eqnarray}
\hbox{Prob}(i|j) &=& \frac{p(i)r(j|i)}{\sum_k p(k)r(j|k)} \\
 &=& \rule{0cm}{1cm} \frac{\tr (G_j \Pi_i)}{d\cdot \tr\, G_j}.
\end{eqnarray}
Now let me redefine this $G_j/ \tr\, G_j$ and call it a density operator.

But then look at this probability.  The probability for getting $i$ in the sky given $j$ on the ground is just one of these SIC representations of the quantum state $\rho_j$:
\begin{equation}
\hbox{Prob}(i|j)= \frac{1}{d} \tr (\rho_j \Pi_i),\ \rho_j
 = \frac{G_j}{\tr\, G_j}.
\end{equation}
So this is generated by a hidden quantum state $\rho_j$.
Moreover, any $\rho_j$ can be gotten in this way by choosing the measurement on the ground to be an appropriate POVM, because that is a property of quantum mechanics:  That for any POVM
we can generate a quantum state this way, and vice versa.

Well, all that was really a statement of was the combination of using Bayes' Rule and using this operator representation.  So, I'm going to promote it to an axiom:
\begin{quote}
Starting from a state of maximal uncertainty for the sky, one can use
the posterior state supplied by Bayes' Rule,
\begin{equation}
\hbox{Prob}(i|j) = \frac{r(j|i)}{\sum_k r(j|k)},
\end{equation}
as a valid prior state.  Moreover all valid priors can be generated in this
way.
\end{quote}
We do this all the time: Whenever you gather data, you take your
prior, you turn that into a posterior, and then you use that later on for your next prior.  So that's all that this rule is telling you, and I have particularized it to having complete ignorance of the sky.

That leads to an immediate consequence for our formula.  Because if we just write it out, now supposing the $r(j|i)$ derive from one of these posteriors, the formula becomes this
\begin{equation}
q(j) = \left(\sum_k r(j|k)\right) \left[(d+1)
 \sum_i p(i)\hbox{Prob}(i|j) - \frac{1}{d}\right].
\end{equation}
But we demand that the number on the left be nonnegative. So that tells us that the
the thing in the square brackets here has to be nonnegative, and by the postulate above,
this Prob$(i|j)$ is just some other quantum state.  And if we want this to be nonnegative, then it says that the inner product of these two quantum states is bounded below.

Within the probability simplex, if I have a point that is a valid state and I have another point that is a valid state---in other words one that doesn't violate the condition here---their inner product can't be too small.  For
any two valid priors $\vec{p}$ and $\vec{s}$,
\begin{equation}
\vec{p}\cdot\vec{s} = \sum_i p(i)s(i) \geq \frac{1}{d(d+1)}.
\end{equation}
That's a nontrivial condition, I would say.

Let me now point out another thing.  If we're going to be consistent---we have the measurement in the sky and we have the measurement on the ground---then it has to be the case that we can do the
calculation for a measurement on the ground that is exactly the same
as the measurement in the sky.  Then, self-consistency requires that for any valid
$\vec{p}$,
\begin{equation}
p(j) = (d+1)\sum_i p(i) r(j|i) - \frac{1}{d}.
\end{equation}
If this is going to hold, then the only conditional probabilities
$r(j|i)$ that can be allowed are ones of this sort:
\begin{equation}
r(j|i) = \frac{1}{d+1}\left(\delta_{ij} + \frac{1}{d}\right).
\end{equation}
Now we use the Reciprocity Axiom, the one about Bayes' Rule, to turn these into actual states: All
$\vec{p}$ of the form
\begin{equation}
\vec{e}_k = \left[\frac{1}{d(d+1)}, \ldots
                  \frac{1}{d}, \ldots,
                  \frac{1}{d(d+1)}
            \right]
\end{equation}
must be valid priors.  It says that among our set of states, just because of the existence of these two measuring devices, we have to
have states which are flat except for one point, with this particular
normalization.  We'll call these \emph{basis distributions.}

Something to notice about these states:  If we take the inner product
of any one of them with itself,
\begin{equation}
\vec{e}_k \cdot \vec{e}_k = \frac{2}{d(d+1)}.
\end{equation}

Since these correspond to our very special measuring device, let's
think of these as being among the extreme points of the set of valid
distributions.  ``Extreme,'' in the sense means that the norm
takes its largest possible value.

This leads to the following notion.  I'll call a set $S$
within the probability simplex $\Delta_{d^2}$, which contains these points $\vec{e}_k$ it must contain, \emph{consistent\/}
if for any two points $\vec{p},\vec{q} \in S$, we have
\begin{equation}
\frac{1}{d(d+1)} \leq \vec{p} \cdot \vec{q}
 \leq \frac{2}{d(d+1)}.
\label{Weedbrook}
\end{equation}
And I will call $S$ \emph{maximal\/} if adding any further point $\vec{p} \in
\Delta_{d^2}$ makes it inconsistent.

So, if I have a set for which every pair of points satisfies Eq.~(\ref{Weedbrook}) and I add just one more point to it from the simplex, if all of a sudden Eq.~(\ref{Weedbrook}) is violated, then I say the set is maximal.

Here's an example of a maximal consistent set: if $S$ is the set of
quantum states itself, it is a consistent and maximal set.  I'll show you
that momentarily.  As a general problem, it would be nice to
characterize \emph{all\/} such sets.  We know that quantum state spaces
are among them, but what else is among them?

Let me show you that quantum state space is a maximal consistent set.
Suppose I use the SIC representation to turn two probability
distributions into operators:
\begin{align}
\rho &= \sum_i\left[(d+1)p(i) - \frac{1}{d}\right]\Pi_i, \\
\sigma &= \sum_i\left[(d+1)q(i) - \frac{1}{d}\right]\Pi_i.
\end{align}
It works out that the Hilbert--Schmidt inner product of these
operators can be written in terms of the normal Euclidean inner
product of the probability distributions:
\begin{equation}
\tr\,\rho\sigma = d(d+1)\vec{p}\cdot\vec{q} - 1.
\end{equation}
Suppose $\vec{p}$ does not correspond to a quantum state.  Well,
$\rho$ is automatically Hermitian, so if it's not going to correspond
to a quantum state, then it has to have a negative eigenvalue.  If
$\rho$ has a negative eigenvalue, let me choose $\sigma$ to be the
projection onto the direction which gives that negative eigenvalue.
I see that I generate a Hilbert--Schmidt inner product that is less
than zero:
\begin{equation}
\tr\,\rho\sigma < 0.
\end{equation}
Consequently, the inner product of probabilities has to be smaller
than our bound:
\begin{equation}
\vec{p}\cdot\vec{q} < \frac{1}{d(d+1)}.
\end{equation}
There's nothing you can add to quantum state space and still satisfy
the consistency condition!

OK. So, quantum state space is in there.  But what other properties do
these maximal consistent sets have in general which are along the lines of
quantum mechanics?

For starters, we can show that maximal consistent sets have to be
convex.  Let $S$ be a consistent set.  If $\vec{p},\vec{q} \in S$,
then for any $\vec{r} \in S$ and $0 \leq x \leq 1$,
\begin{equation}
\frac{1}{d(d+1)}\, \leq \, [x\vec{p} + (1-x)\vec{q}\,]\cdot\vec{r}
\, \leq \, \frac{2}{d(d+1)}.
\end{equation}
Therefore, maximal consistent sets have to be convex sets.

Let $\vec{p}$ belong to the closure of~$S$.  Then there must be a
sequence $\vec{p}_t \in S$ converging to~$\vec{p}$.  But for any
$\vec{q} \in S$,
\begin{equation}
\frac{1}{d(d+1)} \leq \vec{p}_t\cdot\vec{q}
 \leq \frac{2}{d(d+1)}.
\end{equation}
Therefore, if $S$ is a maximal consistent set, then $\vec{p}$ belongs
to~$S$.  So maximal consistent sets are closed.

So we have closed and convex sets just from the two conditions for maximal consistent sets.  In what sense do we get \emph{nontrivial\/} sets?  There's already a
sense in which we get a pretty nontrivial set, just from the lower bound of
the consistency condition.  For any point that's allowed in my set,
there has to be some point on the opposite side of the sphere, around the
antipode, that can't be allowed.  There has to be a nonincluded
region.  For any point, there's an excluded spot on the opposite
side of the sphere, which is pretty strange.

Moreover, we've got this property: The sphere we're talking about
actually reaches outside of the simplex for sufficiently
high-dimensional faces.  Whatever the object is, not only does it have
the property that antipodal regions are excluded, but also it is
larger than the simplex it's sitting in.  So, it has to be the
intersection of a sphere and a simplex.
\begin{center}
\includegraphics[width=9cm]{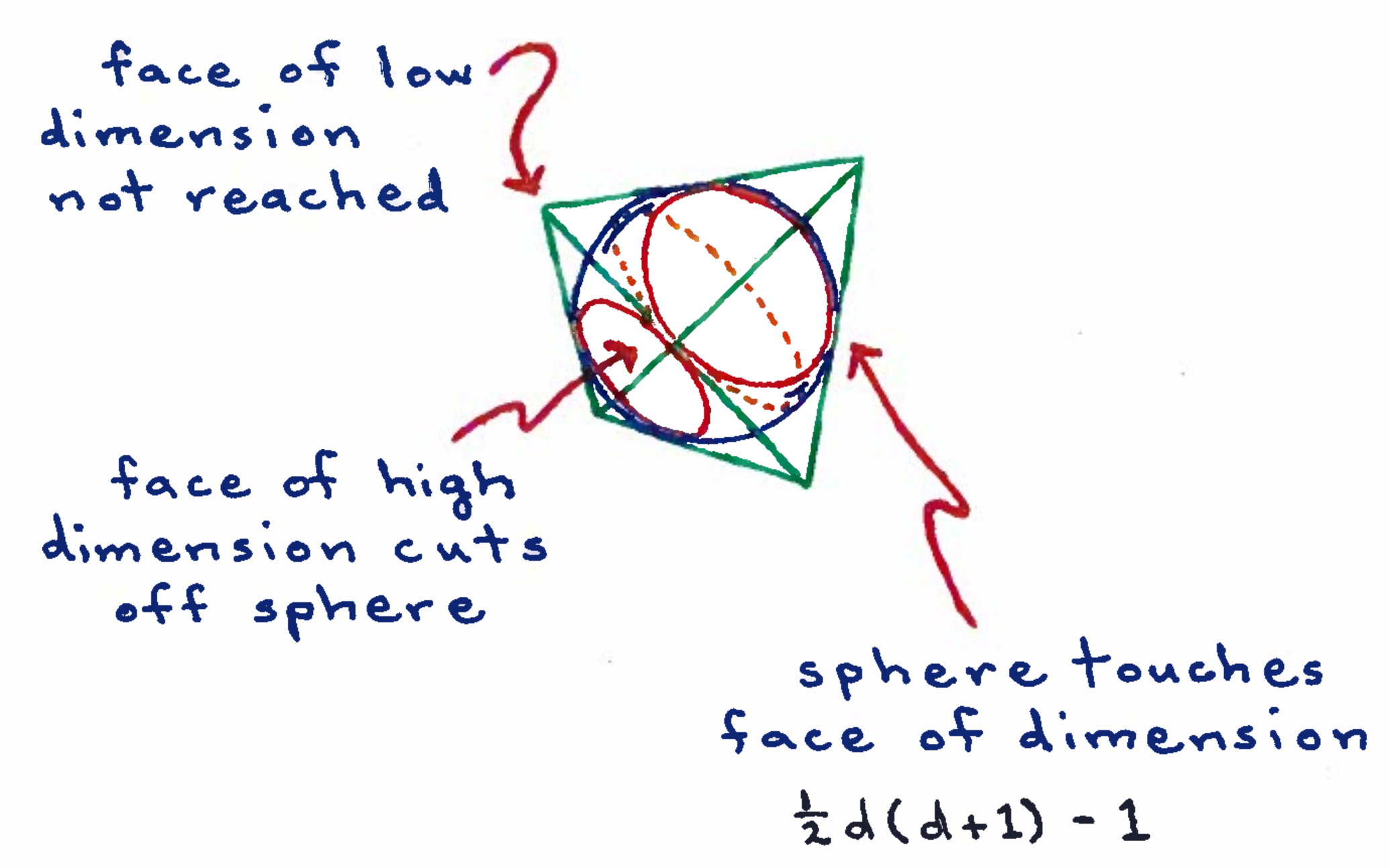}
\end{center}

In more detail, the argument goes like this.  Re-reference points to
the center $\vec{c}$ of the probability simplex:
\begin{equation}
\vec{p}{\,'} = \vec{p} - \vec{c}.
\end{equation}
The consistency condition becomes
\begin{equation}
-\frac{1}{d^2(d+1)} \leq \vec{p}{\,'}\cdot\vec{q}{\,'}
 \leq \frac{d-1}{d^2(d+1)}.
\end{equation}
The sphere is too big for the simplex!

Here's another interesting property.  You might ask yourself, ``Is
there some bound on the number of zeros which can be in a probability
vector?''  You can use the Schwarz inequality to show that indeed
there is.  Suppose $\vec{p}$ has $n$ zero values, $p(i) = 0$.  Then
\begin{equation}
1 = \left(\sum_{p(i)\neq0} p(i)\right)^2
 \leq \, (d^2 - n)\sum_{p(i)\neq0} p(i)^2
\, \leq \, (d^2 - n) \frac{2}{d(d+1)}.
\end{equation}
So,
\begin{equation}
n \leq \frac{1}{2}d(d-1).
\end{equation}
This automatically implies that there are certain asymmetries to this
set.  So it started off in a very symmetric way, but suppose you achieve this bound on the number of zeros:  We have
some $\vec{p}_1$ that has a lot of zeros and then some other values.
\begin{equation}
\vec{p}_1 = (0, 0, 0, 0, p_5, p_6, \ldots, p_n).
\end{equation}
Let's say that this is a good point.  Alternatively, consider another one in which
I've rotated around and moved my zeros one spot to the right:
\begin{equation}
\vec{p}_2 = (p_1, 0, 0, 0, 0, p_6, \ldots, p_n).
\end{equation}
That might still be a good point.  But generally, it'll be the case
that if I take all of my zeros and shuffle them to the other side of the vector,
\begin{equation}
\vec{p}_3 = (p_1, p_2, \ldots, p_k, 0, 0, 0, 0),
\end{equation}
all of a sudden the inner product bound will be violated: $\vec{p}_1 \cdot \vec{p}_3$
is too small.  So it's not the case that all permutations of the vectors' components lead to allowed points within the set.

All of these things are picking up features that the cubic equation in
the quantum state space specification implies.  It's a nice, simple
condition, but it has lots of consequences, and all these things are consequences of
the cubic equation, not the quadratic condition.

Here's one last one.  Suppose we have a set of vectors within one of
these maximally consistent sets, and they're all of the largest length
possible, and moreover, they all have the smallest inner product
allowed by the bound.  That is, we have $\vec{p}_k{'} \in S$, with $k =
1,\ldots,m$, that saturate both bounds:
\begin{align}
\vec{p}_k{'}\cdot\vec{p}_k{'} &= \frac{d-1}{d^2(d+1)}\ \forall k, \\
\vec{p}_k{'}\cdot\vec{p}_l{'} &= -\frac{1}{d^2(d+1)}\ \forall k\neq l.
\end{align}
Here is the question I would like to ask.  For vectors of this variety, is there some maximum number of vectors within this set.  How many can be there?  That is, how large can we make $m$?

You can figure this out by forming this combination
\begin{equation}
\vec{G} = \sum_k \vec{p}_k{'},
\end{equation}
and observing that
\begin{equation}
0 \leq \vec{G}\cdot\vec{G}
 = \sum_{k,l} \vec{p}_k{'}\cdot\vec{p}_l{'}
 = \frac{m(d-m)}{d^2(d+1)}.
\end{equation}
This expression can only be nonnegative if
\begin{equation}
m \leq d.
\end{equation}
So it says we've started with a simplex of size $d^2$, and if we demand that we
have a set of points which are maximally distant from one another, we
can't have more than $d$ of them.  Again, this mimics quantum state
space!  We started with a set of dimension $d^2$ and found a kind of
underlying set of size $d$.

Moreover, if
\begin{equation}
\vec{G}\cdot\vec{G} = 0,
\end{equation}
then
\begin{equation}
\sum_{k=1}^d \frac{1}{d} \vec{p}_k{'} = 0
\ \Rightarrow\ \sum_{k=1}^d \frac{1}{d} \vec{p}_k = \vec{c}.
\end{equation}
This is also the same as in quantum mechanics.

So, challenge:  What further postulates must be made to recover quantum state space
precisely?  That is, how do we recover the convex hull of
\begin{align}
\sum_i p(i)^2 &= \frac{2}{d(d+1)}, \\
\sum_{ijk} c_{ijk} p(i)p(j)p(k) &= \frac{d+7}{(d+1)^3},
\end{align}
with the $c_{ijk}$ possessing the correct properties?  I don't know,
maybe they'll get really horrible.  Maybe it won't be the pretty
principle that I want at all.  But I feel that at least this is moving
in the direction of focusing on contextuality as a primitive notion
within quantum theory, and the right particular flavor of
contextuality.  Of course, what we really want is this thing John Wheeler said:
\begin{quotation}
\noindent If one really understood the central point and its necessity
in the construction of the world, one ought to be able to state it in
one clear, simple sentence.
\end{quotation}
I hope you guys keep trying to do that.  Thank you!

\bigskip\medskip

\noindent {\bf During the Q\&A, there was a comment from Lucien Hardy.} \bigskip

{\bf HARDY:} Maybe first a comment, and then a more specific question.  The way I see it, in terms of this program of finding some postulates or axioms or whatever, is that we're looking at some strange object, and the original mathematical axioms of quantum theory sort of give us a bunch (five or however many there are) of strange vantage points on this strange object.  Maybe we're looking at it from odd directions.  And as we get better postulates, we're finding better ways to look at it from.  So, rather than looking at it from some direction where it doesn't make sense, perhaps we can see it face-on, and we see it from several different directions.  So we're moving towards some more reasonable set of vantage points on the given object, so then it wouldn't disturb us of course whether these axioms are unique or not---at least they should somehow be a sensible way of looking at that object.  Just sort a general comment.  But then the particular point was you said that you thought this equation could be a possible postulate, but you also said that you liked it when postulates were expressed in words.  So do you have some words for that equation? \medskip

{\bf FUCHS:} No, that's my big failure at this point.  But I liked your analogy at the beginning, and I guess maybe it gives me the tool to express that:  It seems that the light has been shining in ways that it kind of takes the corners off of the edge of quantum theory.  So we've got this
sharp, rather jagged object called quantum theory, and the lights that
have been shined on it have made it like a little distinction from
classical theory in usual ways.  For instance in Giulio and Mauro and Paulo's postulates, they
can pinpoint it to one spot, and you can as well.  And some of the other axiom systems.  So it feels to me that it is kind of dulling \ldots\ the projection of
it is smoother than the real object itself is.\footnote{David Mermin says it much better in the paper where he originated the phrase `shut up and calculate'~\cite{Mermin89}:  ``I would rather celebrate the strangeness of quantum theory than deny it, because I believe it still has interesting things to teach us about how certain powerful but flawed verbal and mental tools we once took for granted continue to infect our thinking in subtly hidden ways \ldots.  [T]he problem with the second generation's iron-fistedly soothing attitude is that by striving to make quantum mechanics appear so ordinary, so sedately practical, so benignly humdrum, they deprive us of the stimulus for exploring some very intriguing questions about the limitations in how we think and how we are capable of apprehending the world.''}

But for your question, yeah it's a shortcoming.  I don't have any explanation of this equation that is giving some structure other than to know that it works.  I've tried to play games to do with Dutch book arguments and some kind of picture of the world that \ldots\ uh \ldots\ You know I have these \ldots\ Like Gilles, you see, Gilles was my teacher.  He has conversations with God. So I've had conversations with God where I imagine that God says \ldots\ You know, I ask him, ``I want the ability to write messages upon the world.''  And God says,``You know \ldots\ I can give you that but that means that the world is going to have to have some loose play in it, because if it were a rigid thing then you wouldn't be able to write messages in it.''  Then he says, ``Ah but the moment I give you some loose play in the world, then I may not be able to predict all the floods and fires for you anymore, because it's got all this loose play in it.''  So anyway I've played little games like that associated with Dutch book, and I've gotten nowhere.

\vspace{-0.5cm}
\begin{center}
\includegraphics[width=12cm]{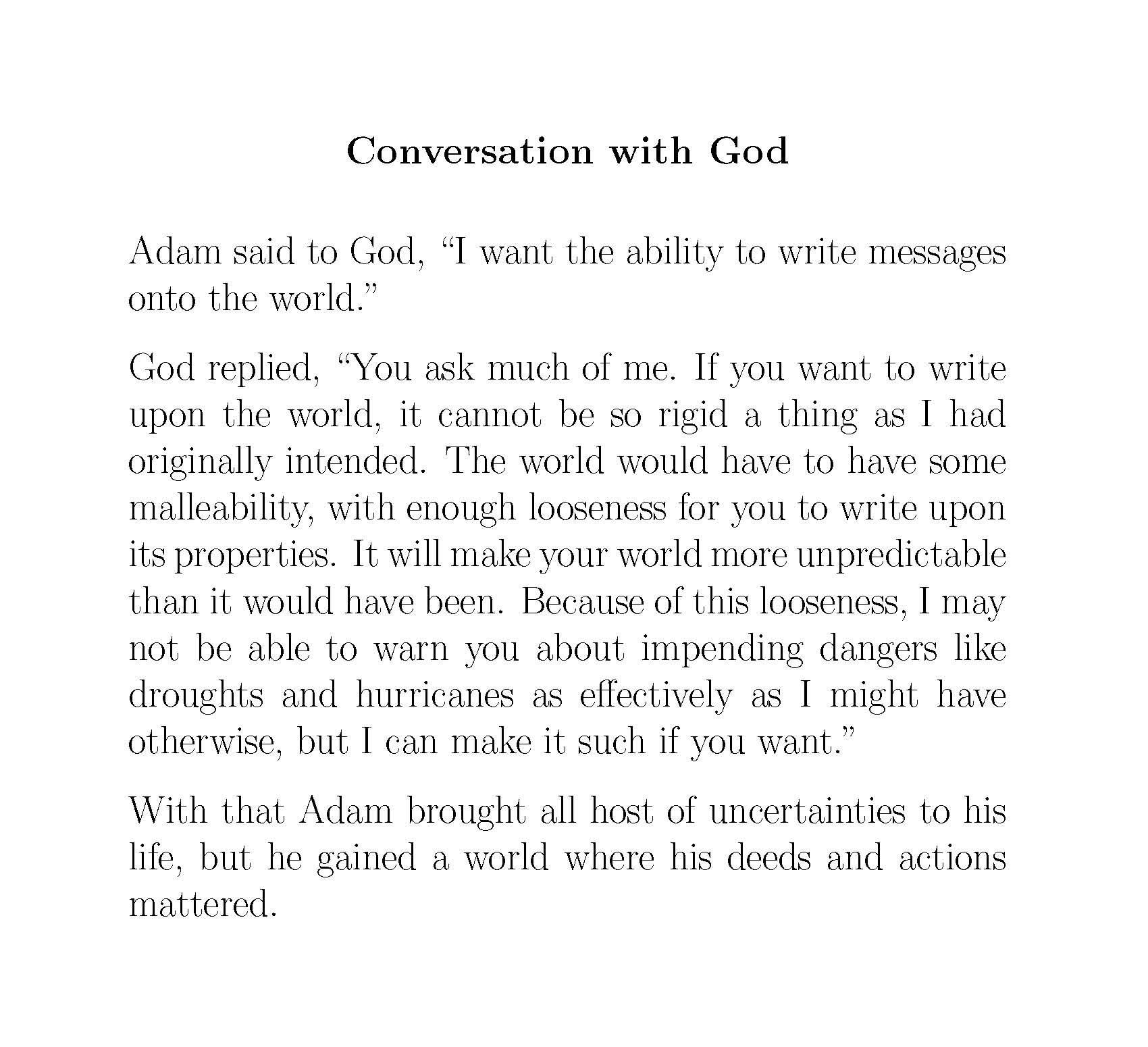}
\end{center}

\medskip

\pagebreak

%

\end{document}